# DYNAMICS OF FEMTOSECOND ELECTRON BUNCHES

A. G. Khachatryan, A. Irman, F. A. van Goor, and K.-J. Boller

Faculty of Science and Technology, MESA+ Institute, University of Twente, P.O. Box 217, 7500 AE Enschede, The Netherlands

In the laser wakefield accelerator (LWFA) a short intense laser pulse, with a duration of the order of a plasma wave period, excites an unusually strong plasma wake wave (laser wakefield). Recent experiments on laser wakefield acceleration [Nature (London) **431**, p.535, p.538, p.541 (2004)] demonstrated generation of ultra-short (with a duration of a few femtoseconds) relativistic electron bunches with relatively low energy spread of the order of a few percent. We have studied the dynamics of such bunches in vacuum and in laser wakefield. The results show strong bunch dynamics already on a few millimeters propagation distance in both cases. In vacuum, the bunch radius and emittance quickly grow. The latter worsens the focusability of the bunch. We found that when a femtosecond bunch is accelerated in a channel-guided laser wakefield, for realistic bunch lengths, the bunch length is approximately conserved. However, the spread in betatron frequencies leads to fast betatron phase mixing in the bunch envelope for on-axis injection. When bunch is injected in a laser wakefield off-axis, the bunch decoherence results in considerable increase in the normalized bunch emittance, and, in some cases, in increase in the energy spread, after acceleration. We also discuss a possible two-stage laser wakefield accelerator.

## I. INTRODUCTION

High-intensity laser pulse with an ultra-short duration of the order of the plasma wave period can generate very strong accelerating and focusing fields (wakefield) in plasma [1, 2]. Extremely large accelerating gradients as high as a few tens of GV/m have been measured in experiments (see, e.g., Ref. [3]), which is three orders of magnitude higher than what can be achieved in conventional accelerators. This makes laser wakefield accelerator (LWFA) very attractive for electron acceleration. The length of the region in the laser wakefield suitable for particle acceleration is less than half of the plasma wavelength, which is typically of the order of a few tens of microns. The transverse extent of the accelerating region, determined by the diameter of the laser pulse, has the same order of magnitude. Then, to have a small energy spread in an accelerated bunch, the bunch sizes must be much less than the plasma wavelength, i.e., a few microns, which corresponds to bunch duration of the order of 10 femtoseconds. Obviously, such bunches cannot be provided by the conventional accelerator technique. The generation of fs electron bunches is, apparently, the main challenge for LWFA. There were some ideas on bunch injection in the laser wakefield, when some plasma electrons are injected in the accelerating phase of the wakefield by using additional laser pulse(s) [4-6], by tunneling ionization and subsequent ponderomotive acceleration [7], or by the wave-breaking mechanism [8-10]. It was also found that injection of an e-bunch longer than the plasma wavelength into a laser wakefield can generate femtosecond relativistic bunches with low energy spread when the injection energy is sufficiently low (typically a few MeVs) [11-19]. Experimental breakthrough came in 2004 when several research groups, working in the bubble regime (sometimes called also blow-out regime), as proposed in [9], demonstrated the generation of fs electron bunches [20-22]. The energy of bunches was of the order of 100 MeV and the relative energy spread was of the order of a few percent. Later, an unprecedented 1 GeV bunch energy and 2.5 % energy spread was obtained by this method when the drive laser pulse was guided in a 33 mm long plasma channel [23]. However, pour shot-to-shot stability was reported. Recently considerable improvement in the stability was demonstrated by the LOA group [24]. This was done by employing a second, counter-propagating laser pulse, which injects some plasma electrons into the accelerating phase of the laser wakefield excited by the first pulse [5]. The measurements and supporting numerical simulations show the following typical bunch parameters in the LWFA experiments: a duration of 10 fs or less, a transverse size of a few microns, a charge of tens of pC, an energy of tens to hundreds of MeVs, an rms energy spread of a few percent, and a normalized emittance of of the order of one micron. These parameters make the femtosecond relativistic electron-bunches a qualitatively new object and also new tool in physical research, mainly due to extremely small bunch sizes. One can expect that such bunches show different dynamics as compared to bunches from standard accelerators.

In this article we study theoretically the dynamics of the fs bunches in a laser wakefield and in vacuum. These are the two situations of main interest, when the bunch is accelerated in a laser wakefield or injected into it, or is transported in a drift space. Our results show that the bunch parameters can change considerably already after a few millimeters propagation in vacuum or in the wakefield. In the following section we consider the dynamics of a single electron in a laser wakefield. In Sec. III the propagation of fs bunches in LWFA is studied analytically, via the envelope equation, and numerically. Sec. IV presents the results of simulations of bunch propagation in vacuum. In Sec. V we study the bunch dynamics in a channel-guided laser wakefield in the case when the bunch is injected off-axis. In the last section we discuss and summarize our results. Some details on calculations and simulations are placed in the appendices.

## II. SINGLE PARTICLE MOTION

Before looking at the dynamics of fs bunches we study the motion of single electrons propagating along the laser wakefield. The equation for the normalized momentum of an electron is (see, e.g., Ref. [12]):

$$\frac{d\mathbf{p}}{d\tau} = -\beta_g (\mathbf{E} + \boldsymbol{\beta} \times \mathbf{B}) - \frac{\nabla a^2}{4\beta_g \gamma}, \qquad (1)$$

where $\mathbf{p}=\boldsymbol{\beta}\gamma$ and $\boldsymbol{\beta}=\mathbf{v}/c$ are the normalized momentum and velocity of the electron respectively, $\mathbf{E}$ and $\mathbf{B}$ are the electric and magnetic fields of the laser wakefield normalized to the non-relativistic wave-breaking field $m_e\omega_p v_g/e$ [2], $\omega_p=(4\pi n_p e^2/m_e)^{1/2}$ being the plasma frequency, $v_g$ is the group velocity of the laser pulse in plasma, $\beta_g=v_g/c$, $\gamma=(1+\mathbf{p}^2+a^2/2)^{1/2}$ is the relativistic factor of the electron. If the laser pulse is guided in a plasma channel, the plasma electron concentration, $n_p$, is taken as its on-axis value. The last term on the right-hand-side of equation (1) represents the ponderomotive force [25], $\mathbf{a}=e\mathbf{A}_0/m_e c$ is the normalized amplitude of the vector potential of the laser field; the laser pulse is linearly polarized. For convenience, we have introduced the dimensionless time $\tau=\omega_p t$. The spatial coordinates are normalized to $1/k_p$, $k_p=\omega_p/v_g$ being the plasma wavenumber. One finds from Eq. (1) that the relativistic factor evolves according to the equation:

$$\frac{d\gamma}{d\tau} = -\beta_g (\boldsymbol{\beta} \cdot \mathbf{E}) - \frac{1}{4\gamma}\frac{\partial a^2}{\partial \xi}, \qquad (2)$$

were $\xi=z-\tau$, $z$ is the normalized longitudinal coordinate. Assuming an axially-symmetrical Gaussian laser pulse with $a=a_0\exp[-(\xi-\xi_c)^2/\sigma^2-r^2/w^2]$, where $\xi_c$ is the centre of the pulse, one finds that, in cylindrical coordinates, only wakefield components $E_r$, $E_z$, and $B_\theta$ are excited [2] and that each field component depends on $\xi$ and $r$. However, because an electron bunch, in general case, does not show axial symmetry (for example, when the bunch is injected off-axis), it is convenient to study the dynamics of electrons in a laser wakefield in Cartesian coordinates. Taking into account Eq. (2), one obtains from (1):

$$\frac{d^2x}{d\tau^2} + \frac{1}{\gamma}\left\{[(x/r)(1-\beta_x^2)-(y/r)\beta_x\beta_y]E_r - \beta_x\beta_z E_z - (x/r)\beta_z B_\theta\right\} - \frac{1}{4\beta_g^2\gamma^2}\left(\beta_x\frac{\partial a^2}{\partial \xi} - \frac{(x/r)}{\beta_g}\frac{\partial a^2}{\partial r}\right) = 0,$$

(3.1)

$$\frac{d^2y}{d\tau^2} + \frac{1}{\gamma}\left\{[(y/r)(1-\beta_y^2)-(x/r)\beta_x\beta_y]E_r - \beta_y\beta_z E_z - (y/r)\beta_z B_\theta\right\} - \frac{1}{4\beta_g^2\gamma^2}\left(\beta_x\frac{\partial a^2}{\partial \xi} - \frac{(y/r)}{\beta_g}\frac{\partial a^2}{\partial r}\right) = 0,$$

(3.2)

$$\frac{d^2\xi}{d\tau^2} + \frac{1}{\gamma}\left\{(1-\beta_z^2)E_z + [(x/r)\beta_x + (y/r)\beta_y](B_\theta - \beta_z E_r)\right\} + \frac{(1-\beta_g\beta_z)}{4\beta_g^2\gamma^2}\frac{\partial a^2}{\partial \xi} = 0,$$

(3.3)

where $x$ and $y$ and $\beta_{x,y,z}$ are the normalized transverse coordinates and the normalized velocity components of the electron correspondingly, $r=(x^2+y^2)^{1/2}$. The velocity components can be found from $\beta_x=\beta_g dx/d\tau$, $\beta_y=\beta_g dy/d\tau$ and $\beta_z=\beta_g(1+d\xi/d\tau)$.

The laser wakefield can be calculated analytically in the case of a uniform plasma and a linear wakefield ($a^2\ll1$) [2, 26]. When laser pulse is guided in a plasma channel or the wakefield is nonlinear ($a^2\sim1$ or larger), one usually has to use a numerical code. The structure and features of channel-guided and nonlinear laser wakefields are described in the literature in detail (see, e.g., [27]). The important fact is that there are regions in the wakefield where the field is both accelerating and focusing. Accordingly, an electron bunch injected in these regions can be accelerated to ultra-relativistic energies while the focusing force keeps the electrons near the wakefield axis. The latter is important to obtain low energy spread in the accelerated bunch because the accelerating field decreases with the distance from the axis and electrons situated relatively far from the axis will gain less energy compared to the on-axis electrons. Therefore, the most practically interesting situation is when electrons are concentrated near the axis. In this case, assuming relativistic electrons co-propagating with the wakefield behind the laser pulse, where $a^2=0$, one finds from (2):

$$d\gamma/d\tau \approx -E_z,$$
$$\gamma \approx \gamma_0 - \int_{\tau_0}^{\tau} E_z d\tau. \tag{4}$$

In (4) $\gamma_0=\gamma(\tau_0)$ is the initial relativistic factor and we assumed that $\beta_g \approx 1$, which is always fulfilled in LWFA. Note also that $E_z<0$ ($E_z>0$) corresponds to electron acceleration (deceleration). Next, without loosing generality, suppose that $y(\tau)=0$. Then, taking into account that for near-axis electrons $E_r \approx (\partial E_r/\partial r)r$, $B_\theta \approx (\partial B_\theta/\partial r)r$, and assuming that $|\beta_x|<<1$, one has from (3.1) [12]:

$$\frac{d^2 x}{d\tau^2} - \frac{E_z}{\gamma}\frac{dx}{d\tau} + \frac{1}{\gamma}\frac{\partial F_r}{\partial r}x = 0, \tag{5}$$

where $F_r=E_r-B_\theta$; $F_r>0$ ($F_r<0$) corresponds to focusing (defocusing) regions of the wakefield. For the following we assume that $E_z$ does not change its sign while the electron stays in the wakefield, i.e., that the electron continuously gains or looses energy. In this case $\gamma$ is a monotonous function of time (if this is not the case, one can divide the electron's trajectory into segments where $\gamma$ is monotonous and consider these segments separately). If the characteristic time scale on which $E_z$ and $\partial F_r/\partial r$ change, which is in the order of the dephasing time $2\pi\gamma_g^2/\omega_p$ (the corresponding dephasing length is $\gamma_g^2\lambda_p$ [2], where $\lambda_p$ is the plasma wavelength), is much larger than the characteristic time scale on which the transverse position of the electron changes (in fact, this means that the betatron period is much less than the dephasing time, which is typically the case) [28], one has the following solution of Eq. (5) (see Appendix A):

$$x = C_1 J_0(s) + C_2 Y_0(s), \tag{6}$$

where $J_0$ and $Y_0$ are the Bessel functions of the first and second kind correspondingly, $s=(b_1\gamma)^{1/2}$, $b_1=4f/E_z^2$, $f=\partial F_r/\partial r$. The constants $C_{1,2}$ can be found from initial conditions: $C_1=-(\pi/2)[s_0 Y_1(s_0)x_0-(2\gamma_0/E_{z0})Y_0(s_0)\beta_{x0}]$, $C_2=(\pi/2)[s_0 J_1(s_0)x_0-(2\gamma_0/E_{z0})J_0(s_0)\beta_{x0}]$, where subscripts "0" denote the initial values. When $f<0$ (defocusing region of the wakefield) the Bessel functions in (6) are transformed to the modified Bessel functions which have monotonous behavior. This corresponds to ejection of an electron from the wakefield. When $\beta_{x0}=0$ and $s>>1$, one has from (6):

$$x = x_0(s_0/s)^{1/2}\cos(s-s_0). \tag{7}$$

In the focusing region, where $f>0$, expressions (6) and (7) describe so-called betatron oscillations of the electron near the axis. The amplitude and frequency of the oscillations are determined by the wakefield. The betatron frequency can be determined as $\omega_\beta=ds/dt=\omega_p(f/\gamma)^{1/2}$ (see also Refs. [12, 29]). It follows from (7) that the amplitude of the betatron oscillations is proportional to $(\gamma_0/\gamma)^{1/4}$, the betatron frequency scales as $\gamma^{1/2}$ and $\beta_x \sim 1/\gamma^{3/4}$. One can neglect the phase slippage when the laser pulse travels in plasma over a distance much less than the dephasing length. In this case $\xi$, $E_z$ and $f$ do not change much in time and $s \sim \gamma^{1/2} \sim \tau^{1/2}$.

To study the electron dynamics one needs to describe the laser wakefield. In general, the values of $f$ and $E_z$, which are involved in Eqs. (4)-(7), are related to each other via $f=\partial^2(\int E_z d\xi)/\partial r^2$ (see, e.g., Ref. [2]). The linear wakefield in a uniform plasma generated by a linearly-polarized Gaussian pulse with $a=a_0\exp(-r^2/w^2-\xi^2/\sigma^2)$, can be described by $E_z=-E_0\exp(-2r^2/w^2)\cos(\xi)$ [2], where $E_0 \sim a_0^2$ is the amplitude of the field, which for the optimum case of $\sigma=2$ is $\approx 0.4 a_0^2$, constants $w$ and $\sigma$ determine the pulse sizes. In this case, for near axis electrons with $(2r)^2 \ll w^2$, one can use the following approximation:

$$\begin{aligned} E_z &\approx -E_0 \cos(\xi), \\ f &\approx E_0 (2/w)^2 \sin(\xi). \end{aligned} \quad (8)$$

In the case of a wide plasma channel ($w \gg 1$) there is an approximate analytical solution for linear laser wakefield [27]. However, in general there is no analytical solution for the laser wakefield and it needs to be calculated numerically. As a typical example, in Figs. 1 and 2 we show the calculated components of a channel-guided laser wakefield generated by the Gaussian pulse with $w=5$, $\sigma=2$ and $a_0=0.9$. The unperturbed plasma electron density is given by the usual parabolic profile $n_p(r)/n_p(r=0)=1+4r^2/w^4$ [2] (remember that $r$ and $w$ are normalized here). The wakefield was calculated by our fluid-Maxwell code [12, 30] and we will use the shown wakefield to study the dynamics of electrons and e-bunches in such laser wakefield.

In Fig. 3 we demonstrate the transverse dynamics of four electrons in the channel-guided laser wakefield presented in Figs. 1 and 2 as a function of the laser pulse propagation distance $L_{prop}$. The initial gamma factor, $\gamma_0$, is equal to 200 (energy of about 100 MeV) and the initial transverse momentum is zero for all electrons. The electrons are injected with the same initial normalized transverse position $x_0=1$ but with different initial longitudinal positions, $\xi_0=-9.5, -10.5, -11.5,$ and $-13.5$ (see Figs. 1 and 2). The electrons' motion was calculated numerically from Eqs. (3). We chose the plasma wavelength as $\lambda_p$=60 μm, which corresponds to an on-axis plasma electron concentration of $\approx 3\times 10^{17}$ cm$^{-3}$. The spot-size-corrected gamma factor corresponding to the laser group velocity in plasma [2] is $\gamma_g$=70, the laser wavelength is $\lambda_L$=800 nm, the plasma channel length is 5 cm, which is shorter than the dephasing length of $\approx$29 cm in this case. Electrons injected into the defocusing-accelerating and defocusing-decelerating regions, correspondingly at $\xi_0$=−10.5 and −11.5, are deflected by the wakefield and leave the interaction region after a couple of millimeters of propagation in plasma. These cases correspond to the two monotonously growing and closely spaced curves in Fig. 3. The same dynamics was observed in a defocusing region for much smaller initial off-axis positions. The electron injected in focusing-accelerating region ($F_r$>0 and $E_z$<0) at $\xi_0$=−9.5 (see the curve with longer "wavelength") performs a typical betatron oscillation with its amplitude and wavelength decreasing due to the monotonously increasing energy of the electron, which agrees well with expressions (6) and (7). At the end of the plasma channel the gamma factor for this electron reaches a value of about 1500, which corresponds to energy of $\approx$0.77 GeV. The electron injected in decelerating-focusing region ($F_r$>0 and $E_z$>0) with $\xi_0$=−13.5 (the curve with shorter "wavelength" in Fig. 3) is kept in the wakefield but looses its energy until $L_{prop}/\lambda_p \approx$300 (corresponding to $\approx$1.8 cm) where it reaches its minimum energy, $\gamma$=27. Thereafter it is accelerated to $\gamma$=344 at the end of the plasma channel. It can clearly be seen that the amplitude and the wavelength of the betatron oscillations are mainly determined by the energy of the electron: the larger the energy, the longer the wavelength and the lower the oscillation amplitude, as is predicted by expressions (6) and (7). The results show strong correlation between the injection energy, the injection phase and the electron's dynamics. The equations and expressions derived in this section are used below for studying the dynamics of a bunch of electrons.

### III. THE BUNCH ENVELOPE DYNAMICS

In an experiment one deals not with a single electron but with a bunch of electrons. Consider an axially symmetric e-bunch injected in a laser wakefield on-axis. Then one obtains an equation for the bunch radius (envelope equation) by adding to Eq. (5) a term associated with the radial pressure due to a finite bunch emittance (see, for example, [31]):

$$\frac{d^2\sigma}{d\tau^2} - \frac{E_z}{\gamma}\frac{d\sigma}{d\tau} + \frac{f}{\gamma}\sigma - \frac{(k_p\varepsilon_n)^2}{\gamma^2\sigma^3} = 0. \qquad (9)$$

Here $\sigma$ stands for the normalized root-mean-square (rms) bunch sizes $\sigma_x$ and $\sigma_y$, $\varepsilon_n$ is the corresponding normalized rms emittance and $\gamma$ is the mean relativistic factor of the bunch. Eq. (9) actually assumes that all electrons in the bunch experience the same $E_z$ and $f$, which are, in general, functions of time. When the dependence $\gamma(\tau)$ is monotonous and $E_z$ and $f$ are "slow" functions of time, as we discussed in the previous section, one can find the general solution of Eq. (9) (see Appendix B for details):

$$\sigma^2 = A_1 J_0^{\,2}(s) + A_2 J_0(s) Y_0(s) + A_3 Y_0^{\,2}(s). \tag{10}$$

The constants $A_{1,2,3}$ can be found from initial conditions and from equation $A_2^2 - 4 A_1 A_3 = \pi^2 b_2$, where $b_2 = (2 k_p \varepsilon_n / E_z)^2$. When $(2s)^2 \gg 1$ (which corresponds to $\gamma \gg E_z^2/16f$), assuming that the wakefield is focusing, one finds (see Appendix B):

$$\sigma^2 = \frac{C}{2 b_1 \gamma^{1/2}} \left\{ 1 + \Delta \sin[\pm 2s(1 - (\gamma_0/\gamma)^{1/2} + D)] \right\}, \tag{11}$$

where $C$ is the initial value of the function $b_1 \gamma^{1/2} \sigma^2 + b_2/\gamma^{1/2} \sigma^2 - (2\gamma^{1/2}\sigma/E_z)(d\sigma/d\tau) + (4\gamma^{3/2}/E_z^2)(d\sigma/d\tau)^2$, $\Delta = (1 - 4 b_1 b_2 / C^2)^{1/2}$, $D = \arcsin[\Delta(1 - 2 b_1 \gamma^{1/2} \sigma_0^2 / C)]$, the $\pm$ sign in (11) corresponds to the sign of the function $\sigma - (4\gamma/E_z)(d\sigma/d\tau)$. One can see from (11) that the bunch radius oscillates with twice of the betatron frequency for a single electron, which is well-known. From (11) it follows also that the transverse size of the bunch oscillates between some maximum and minimum values given by

$$\sigma_{\text{max,min}} = \gamma^{-1/4} [C(1 \pm \Delta) / 2 b_1]^{1/2}. \tag{12}$$

To further simplify Eqs. (11) and (12) suppose that initially $d\sigma/d\tau = 0$ and that initial bunch energy is sufficiently large, so that $C \approx b_1(\tau_0) \gamma_0^{1/2} \sigma_0^2$. Then, if additionally $4 b_1 b_2 / C^2 \ll 1$, one has:

$$\sigma_{\text{max,min}} = \gamma^{-1/4} \begin{cases} (C/b_1)^{1/2}, \\ (b_2/C)^{1/2}. \end{cases} \tag{13}$$

In this case the maximum bunch radius is determined by the focusing gradient of the wakefield and the minimum radius is given by the bunch emittance. According to Eqs. (10)–(13) the dynamics of the bunch envelope in a laser wakefield, in general, can be quite complicated, which requires a numerical analysis. In the most interesting case of bunch propagation in the focusing phase of the wakefield, the bunch radius evolves as $\sim 1/\gamma^{1/4}$ and performs oscillations determined by the strength of the focusing field and the bunch emittance. From Eq. (10) one can find the bunch radius matched to the laser wakefield taking $d\sigma/d\tau=0$:

$$\sigma_{match} = (k_p \varepsilon_n)^{1/2} / (f\gamma)^{1/4}, \qquad (14)$$

or, in dimensional units,

$$\sigma_{match} = (\lambda_p \varepsilon_n / 2\pi)^{1/2} / (f\gamma)^{1/4}. \qquad (15)$$

One should, however, remember that the bunch radius is not a perfect constant but slowly changes due to the change in the focusing gradient $f$ and the bunch energy.

To study the bunch dynamics numerically we solved equations (3) for an e-bunch with Gaussian distribution of density in both the longitudinal and transverse directions. The details of the calculations are given in Appendix C. The bunch is injected in the wakefield presented in Figs. 1 and 2 at $\xi=-9.5$ with normalized emittances $\varepsilon_{nx}$ and $\varepsilon_{ny}$ of 0.95 µm. The plasma wavelength is taken as 60 micron, the initial bunch energy is 204.4 MeV ($<\gamma>_0=400$), and the initial rms energy spread in the bunch is 2 %. For now, to avoid the effect of finite bunch duration, which we discuss below, the initial bunch length was chosen to be rather short, with a full-width-at-half-maximum (FWHM) bunch length of $3.75\times10^{-3}\lambda_p$ (or 0.225 µm). In Fig. 4 we show the dynamics of the rms bunch radius during acceleration in a 5 cm long plasma channel for initial bunch sizes $\sigma_0$ of 0.96, 1.91, and 2.87 µm. Large variations in the bunch radius can be seen for $\sigma_0=0.96$ and 2.87 µm, while the case of $\sigma_0=1.91$ µm, which is closer to the matching value of $\sigma_{match}\approx2$ µm, according to (15), indeed shows relatively small variations in the radius. In all cases the bunch energy grows monotonously to $\approx850$ Mev while the rms energy spread decreases from 2% to $\approx0.5$ % and the bunch length is slightly increased. The normalized emittances $\varepsilon_{nx,y}$ grow by about 15% and 35% for $\sigma_0=0.96$ µm and $\sigma_0=2.87$ µm, respectively, and are essentially constant for $\sigma_0=1.91$ µm. When the bunch is injected in the laser wakefield with the same emittance but with a smaller radius, namely $\sigma_0<0.5$ µm, some electrons, which have sufficiently large transverse momentum due to the finite emittance, escape the wakefield in the transverse direction. One may call this phenomenon bunch "evaporation". The same happens when the initial bunch radius is too large ($\sigma_0>9$ µm). In this case the focusing force at large transverse positions is not strong enough to keep electrons with a relatively large transverse momentum in the wakefield. The bunch emittance remains constant in a perfectly linear focusing field. This is the case when the bunch stays close to the axis of the wakefield. When the bunch radius becomes larger (which is the case for $\sigma_0=0.96$ µm and $\sigma_0=2.87$ µm), such that the bunch sees the nonlinearity of the focusing field, the emittance grows. We also observed in our simulations that an e-bunch injected in the decelerating–focusing region is first decelerated to $\gamma < \gamma_g$, then slips backwards relative to the wakefield and can thereafter be trapped and re-accelerated in the accelerating–focusing region.

When longer bunches are accelerated in LWFA, the different parts of the bunch with different axial positions in the field, will perform betatron oscillations with different phases due to the different initial conditions (different $f_0$ and $E_{z0}$). This process can be called betatron phase mixing. After propagating some distance in plasma, betatron phase mixing length, different parts of the bunch will even show opposite betatron phases. To simulate more realistic case of a longer bunch we chose a bunch 10 times longer than in the previous simulations presented in Fig. 4, i.e., a FWHM bunch length of 2.25 μm corresponding to a duration of 7.5 fs. $\sigma_0$=2.87 μm, the other parameters were kept the same as used for Fig. 4. In Fig. 5 we show, in the $x$–$z$ plane, the injected bunch (a) and the same bunch after acceleration in a 5 cm plasma channel (b). One can see that the bunch radius is not constant along the bunch length. In an animation of the transition from (a) to (b) we clearly saw the betatron phase mixing in the bunch. The normalized rms emittance grows by 50 % in this case. For typical duration of a bunch produced in bubble regime, i.e., of the order or less than 10 fs, the phase mixing length is of the order of a few millimeters. The phase mixing can be analyzed by considering the evolution of the betatron phase $s$ (see Eqs. (6) and (7)) the same way as it is done in Section V. When the bunch radius is calculated for all electrons in the bunch, it shows typically quickly damped betatron oscillations. This is demonstrated in Fig. 6 for the same initial bunch parameters as in Fig. 5 (compare with the corresponding curve for a shorter bunch in Fig. 4, where the phase mixing can be neglected).

## IV. BUNCH DYNAMICS IN VACUUM

In this section we study propagation of a femtosecond e-bunch in vacuum. This is important when one considers staging of laser wakefield accelerator [32-35], with drift regions between the stages, or when the bunch is to be transported to a target.

Outside plasma, in vacuum, where $E_z=f=0$, one finds from Eq. (9) that the bunch radius evolves according to the well-known expression [31]

$$\sigma = \sigma_0 (1 + z^2 / Z_b^2)^{1/2}, \qquad (16)$$

where, in dimensional units, $Z_b = \gamma \sigma_0^2 / \varepsilon_n$ is the characteristic distance on which the bunch radius grows, $z$ is the propagation coordinate, and $\sigma_0$ is the bunch radius at focus, where $z=0$. At large distances, when $z^2 >> Z_b^2$, one has: $\sigma = \sigma_0 z / Z_b$. Eq. (16) shows that for a typical electron bunch generated by a laser wakefield, $Z_b$ is only a few millimeters. For example, when the normalized emittance is 1 µm, the bunch energy is 200 MeV, and $\sigma_{x0} = \sigma_{y0} = 2$ µm (corresponding to a FWHM transverse bunch size of $2.355\sigma_0 = 4.71$ µm), the value of $Z_b$ is only 1.57 mm. Such rapid grows of the bunch radius during propagation in vacuum is one of the main features of e-bunches produced in LWFA experiments, along with their small sizes. This makes preserving the bunch radius during transportation in vacuum rather difficult. To reveal more details we have simulated the propagation of a fs-bunch in vacuum based on equations (3). The bunch parameters were chosen as $\sigma_{x0} = \sigma_{y0} = 1.91$ µm (the focal value), $<\gamma> = 400$, $\varepsilon_{nx} = \varepsilon_{ny} = 0.95$ µm, FWHM duration of 7.5 fs, and an rms energy spread of 2 %. In Fig. 7 we show the bunch, in the $x-z$ plane, after 50 cm propagation. Upon propagation, except at short initial distance of the order of $Z_b$, the bunch radius was found to grow linearly to a size of $\sigma_{x,y} \approx 617$ µm at a propagation distance of 50 cm, in good agreement with expression (16). This is more than 300 times larger than the initial value. In addition, the bunch becomes 25 % longer while the emittance grows approximately by a factor of 6.5 (see Fig. 8). The latter is the feature of the rms normalized emittance [36], contrary to the behavior of so-called trace-space emittance which remains constant in vacuum (see also Appendix C). Furthermore we have studied the effect of the energy spread on the emittance for the same propagation distance and initial bunch parameters as we used in Figs. 7 and 8, except for the rms energy spread, which was varied in the range from 1% to 10%. We found that the normalized emittance after bunch propagation in vacuum is approximately proportional to the relative energy spread. The bunch lengthening in vacuum can be attributed to both the finite energy spread and the divergence in the initial bunch as follows. When the energy spread is not zero, faster particles travel a longer distance compared to slower particles. This leads to bunch lengthening. Electrons propagating under an angle to the propagation axis, due to the finite divergence (emittance), end up with a larger transverse position but also with a smaller $z$ compared to on-axis particles; this effect also contributes to the lengthening and can be seen in Fig. 7. The bunch lengthening $\Delta l$ due to these two effects can be estimated from expressions $\Delta l \sim \delta \gamma L_{prop} / \gamma^2$ and $\Delta l \sim \sigma^2 / 2 L_{prop} = L_{prop} \varepsilon_n^2 / 2 \gamma^2 \sigma_0^2$ respectively, where $\delta \gamma$ is the relative energy spread. These formulas agree well with the lengthening observed in the simulations.

Another phenomenon which affects the bunch parameters during propagation in vacuum and was not included in the previous simulations is the interaction between bunch electrons (space charge effect). To study the effect of the space charge on femtosecond bunch propagation in vacuum we have used the GPT code [37] for different energies and charges of the bunch. For the calculations we choose the same initial duration, transverse sizes and rms energy spread as in Figs. 7 and 8; the bunch propagation distance is again fixed to 50 cm. The results of the simulations are presented in Table I. Surprisingly, the space charge has minor effect on the bunch parameters during propagation in vacuum when bunch charges of a few tens of pico-Coulomb are considered, which is typical for LWFA experiments [20-24]. This can be explained as follows. In such bunches initial radius is rather small (a few microns) and for a typical normalized rms emittance of the order of a micron the initial transverse momentum of electrons is already relatively large. Therefore, for the considered propagation distance and charge of the bunch, the transverse momentum caused by the space charge has only small effect on the bunch parameters. According to Table I, an appreciable change in the parameters of the bunch can be expected only for bunch charges larger than 100 pC. This concerns mainly the emittance of the bunch while the bunch length after propagation does not much depend on the bunch charge; we found an rms bunch length of 3.2, 1.2 and 0.98 microns, compared with the initial value of 0.955 μm, for $<\gamma>_0$=200, 400 and 800 respectively. This shows that the energy spread and the divergence (emittance) play the dominant role in the bunch lengthening as discussed earlier in this section.

V. BUNCH DYNAMICS IN LASER WAKEFIELD: OFF-AXIS INJECTION

Finally, we consider the propagation of a relativistic femtosecond e-bunch, which is not centered at the wakefield axis and may oscillate as whole around the axis. This takes place when the bunch is injected into a wakefield off-axis, e.g., in a multi-stage laser wakefield accelerator [32-35]. Off-axis propagation can also occur when the laser pulse is guided along a slightly curved trajectory. For example, when the plasma waveguide is slightly bent [38] the laser pulse would follow the curved waveguide while the accelerated relativistic e-bunch would tend to fly in a straight line and to become located off-axis. On the other hand, to preserve the quality of the bunch in a laser wakefield it should be injected sufficiently close to the wakefield axis, where the transverse field is linear. In an experiment his means that, for a typical channel-guided laser wakefield, it is desirable that the off-axis displacement of the bunch is held below approximately 10 microns.

Obviously, when an electron bunch is injected off-axis into the focusing region of a laser wakefield, the bunch, as whole, will tend to oscillate around the axis and perform betatron oscillations like a single electron. The latter is correct if one can neglect the finite bunch size and its longitudinal and transverse emittances. In a real bunch, however, different electrons will oscillate in the wakefield with different betatron phases (frequencies) depending on their initial position and momentum. Even for a high-quality bunch, with sizes much smaller than the plasma wavelength, this can lead to so-called bunch decoherence when, after some propagation distance, the difference in betatron phases of the electrons becomes so large that they occupy the entire region between maximum and minimum displacement.

One can analyze the decoherence analytically as follows. Consider two relativistic electrons injected in the focusing–accelerating phase of a laser wakefield at axial positions $\xi_0$ and $\xi_0+\delta\xi$ and having initially the same betatron phase. According to (6) and (7), the betatron phase can be taken as $\varphi_\beta = s = 2(f\gamma)^{1/2}/|E_z|$, being a function of time and the initial axial position. Next, suppose that the injected bunch length $l_b$ is much smaller than the plasma wavelength, that is, in normalized units, $l_b \ll 1$. Taking into account that the bunch length (and, therefore, also $\delta\xi$) is conserved during bunch propagation in the field with high accuracy, one obtains the difference in betatron phases as $\delta\varphi_\beta = (\partial\varphi_\beta/\partial\xi_0)\delta\xi$. Assuming that the near-axis laser wakefield is described by (8), and that $\gamma^2 \gg \gamma_g^2$ one has: $\xi \approx \xi_0 + \tau/2\gamma_g^2$, $\gamma = \gamma_0 + 2\gamma_g^2 E_0[\sin(\xi) - \sin(\xi_0)]$. Note that the value of $\xi$ in these expressions differs from what is shown in Figs. 1 and 2 by a constant; the focusing–accelerating phase corresponds to $0 < \xi < \pi/2$ in this case. Then one obtains:

$$\delta\varphi_\beta = \frac{4l_b\gamma^{1/2}}{wE_0^{1/2}}\left[Q(\xi) - \left(\frac{\gamma_0}{\gamma}\right)^{1/2} Q(\xi_0) - \frac{\gamma_g^2 E_0}{\gamma}\left(\frac{\cos(\xi_0)}{\cos(\xi)} - 1\right)\sin^{1/2}(\xi)\right], \qquad (17)$$

where $Q(\xi) = [1/2 + \tan^{1/2}(\xi)]/\sin^{1/2}(\xi)$ and we choose the longitudinal distance between electrons to be equal to the bunch length. By taking $|\delta\varphi_\beta| = \pi$ in (17) one can find the appropriate decoherence time $\tau_{dec}$, at which the electrons have moved into opposite betatron phases. The analysis of expression (17) shows that the bunch decoherence becomes stronger with the propagation distance, for larger $\gamma_g$, for larger injection phase $\xi_0$ and for smaller $\gamma_0$. The value of $\delta\varphi_\beta$ quickly grows when $\xi$ approaches the edge of the focusing–accelerating region, where $\xi=\pi/2$. Stronger accelerating field also leads to stronger bunch decoherence. According to (8) and (17), choosing a broader laser pulse (larger $w$) one can weaken the focusing gradient and the process of bunch decoherence. However, this requires more powerful laser. Note also that, according to (11), the phase mixing process considered in Section III proceeds twice as fast as the process of bunch decoherence.

To study the evolution of an e-bunch in the case of off-axis injection in more detail, we simulated the dynamics of the bunch in the wakefield shown in Figs. 1 and 2 for the following initial bunch parameters: $<\gamma>_0=400$, $\varepsilon_{x,y}=0.95$ μm, $\sigma_{x,y}=1.43$ μm, rms energy spread of 2%, FWHM bunch duration of 7.5 fs, with the bunch center located at $<x>_0=9.55$ μm and $<y>_0=0$, the plasma channel length is again 5 cm. Four runs were made for different longitudinal injection positions, where the bunch was injected at phases of 0, 0.3, 0.6, and 0.9 rad from the maximum accelerating field (this corresponds to $\xi$ in expression (8)), which corresponds to $\xi=-9.6, -9.3, -9, -8.7$ in Figs. 1 and 2. The results are shown in Fig. 9. One can see that the bunch center performs damped oscillations determined by the energy of the bunch as expected. The final bunch energy is 880, 823, 728 and 608 MeV for injection phases of 0, 0.3, 0.6, and 0.9 rad, respectively. Fig. 9(b) shows a considerable grows in $\sigma_x$ during acceleration, which witnesses a strong bunch decoherence. Once the process of decoherence completed, the bunch radius approaches to some saturation value, which, for the parameters of Fig. 9, is about 4.5 μm for all injection phases. This is about three times larger than the initial bunch radius. In the $y$-direction the bunch radius shows the same behavior as in the case of on-axis injection, shown in Fig. 6. The most dramatic change experiences the normalized emittance $\varepsilon_{nx}$, it grows more than a factor of 10, $\varepsilon_{ny}$ grows typically by 30-50 %. Accordingly, the quality of the accelerated bunch worsens appreciably. The relative energy spread after acceleration decreases for zero injection phase but is larger than the initial value of 2 % for larger injection phases. From Figs. 9(c) and 9(d) one can conclude that injection at zero phase provides better bunch quality, i.e., smaller emittance and energy spread. In Fig. 10, the electron distribution in the accelerated bunch is depicted in the $x$–$z$ plane for the case of zero injection phase. The bunch shows a typical "snake-shaped" distribution, which is caused by the fact that electrons with different longitudinal positions have different betatron phases (frequencies), as we discussed above. We have observed the same bunch shape in the $p_x$–$z$ plane. Our simulations for an e-bunch which is 10 times shorter (the corresponding FWHM duration of 0.75 fs) show that the bunch decoherence is weaker at the same propagation distance, in agreement with the results on betatron mixing we presented in Section III. These results suggest that the bunch length has to be two orders of magnitude shorted than the plasma wavelength if one needs to avoid the betatron phase mixing and the bunch decoherence. This may experimentally be achieved by increasing the plasma wavelength to a few hundreds of microns. So, the results show that off-axis bunch injection into a laser wakefield may have dramatic effects on the bunch quality.

## VI. SUMMARY AND DISCUSSION

We have studied the dynamics of the recently realized unprecedented micron-sized relativistic-electron-bunches when propagating through vacuum and along with a laser wakefield. Our results show that strong bunch dynamics are expected already upon a few millimeters propagation distance in both cases. When such a bunch of realistic length is accelerated in a channel-guided laser wakefield, the bunch length is approximately conserved. However, the spread in betatron frequencies leads to fast betatron phase mixing in the bunch envelope for on-axis injection. When the bunch is off-axis injected into a laser wakefield, the bunch decoherence results in a considerable increase in bunch emittance, and, in some cases, in an increase in the energy spread, after acceleration. Ironically, a femtosecond bunch is not short enough to avoid bunch decoherence and the accompanying degradation of the bunch quality. Our results suggest that, to avoid the degradation, the spatial extent of the accelerating region is to made larger than that in typical laser wakefield. A solution might be to increase the plasma wavelength towards a few hundreds of microns. This would have another advantage, namely, a longer dephasing length. Another important result is that the transverse size of an e-bunch injected into a laser wakefield is to be matched to the focusing gradient. Otherwise, the bunch radius would oscillate over a wider range during acceleration in the wakefield. This would then degrade the emittance due to the nonlinearity of the focusing field at large distances from the axis.

When propagating in vacuum, the radius of a fs-bunch quickly grows, on a characteristic propagation distance of couple of millimeters. On the other hand, drift space with length of a few tens of centimeters may be needed between LWFA stages to place different elements, for example, a focusing mirror for the drive laser pulse for the next acceleration stage. After such propagation distances the bunch radius can grow by a factor of several hundreds. To inject the bunch into the next stage, it needs to be refocused to about the initial radius of a few microns. This seems to be problematic with the use of conventional focusing elements, like quadrupole magnets. Furthermore, the emittance of the bunch can considerably grow during propagation in the drift space, which means that the focusability of the bunch degrades. Therefore, it is desirable to make the distance between the accelerator stages as short as it is possible. Taking this into account, one can consider the following two-stage laser wakefield accelerator, which is schematically depicted in Fig. 11. A first laser pulse (shown by solid arrows) is tightly focused to a few micron radius to a gas jet (the gas jet extents in the laser propagation direction typically by one or two millimeters; it is depicted in Fig. 11 by the grey ellipse) and produces a relativistic femtosecond bunch in the bubble regime. The femtosecond e-bunch (shown by the black ellipse) from this fist acceleration stage is injected into the channel-guided laser wakefield driven by a second laser pulse (depicted by dashed arrows in Fig. 11), which is focused to a spot radius typical for this regime, namely, a few tens of microns. Because the first laser pulse is tightly focused it diverges quickly; its Rayleigh length is less than one millimeter for the typical laser wavelength of 800 nm. If the distance between the accelerating stages is of the order of a few millimeters, the intensity of the first pulse at the entrance of the plasma channel can be made much less than the intensity of the second laser pulse, such that the first pulse has only a minor influence on the wakefield in the channel. The e-bunch from the first accelerating stage is further accelerated in the plasma channel where the plasma wavelength is essentially larger than that in the gas jet. This can considerably improve the relative energy spread in the bunch if $\lambda_p$ is sufficiently large in the channel. The timing between the two laser pulses can be relatively easily controlled if the pulses are derived from the same laser system.

## ACKNOWLEDGMENTS

This work has been supported by the Dutch Ministry of Education, Culture and Science (OC&W), by Dutch Foundation for Fundamental Research on Matter (FOM) under the "Laser Wakefield Accelerators" program, and by the EuroLEAP project.

## APPENDICES

### A: Solution of equation (5)

With substitution $\theta = \gamma^{1/2}$, taking into account (4), one has from (5):

$$\frac{d^2x}{d\theta^2} + \frac{1}{\theta}\frac{dx}{d\theta} + \left(\frac{4}{E_z^2}\frac{\partial F_r}{\partial r}\right)x = 0. \tag{A1}$$

Then, introducing a new variable $s=[4(\partial F_r/\partial r)/E_z^2]^{1/2}\theta$ one obtains the Bessel equation from (A1), the general solution of which is obviously given by expression (6).

### B: Solution of equation (9)

Again, introducing $\theta=\gamma^{1/2}$ one finds from (9) the following equation for $\rho=\theta^{1/2}\sigma$:

$$\frac{d^2\rho}{d\theta^2} + \left(\frac{1}{4\theta^2}+b_1\right)\rho - \frac{b_2}{\rho^3} = 0, \tag{B1}$$

where $b_1=4f/E_z^2$ and $b_2=(2k_p\varepsilon_n/E_z)^2$. One can show that the general solution of Eq. (B1) is [39, 40]:

$$\rho^2 = A_1 g_1^2 + A_2 g_1 g_2 + A_3 g_2^2, \tag{B2}$$

where $g_1$ and $g_2$ form a fundamental set of solutions of the auxiliary differential equation

$$\frac{d^2 g}{d\theta^2} + \left(\frac{1}{4\theta^2}+b_1\right)g = 0, \tag{B3}$$

the constants $A_{1,2,3}$ satisfy the initial conditions and also the equation $A_2^2-4A_1A_3=4b_2/W^2$, where $W=g_1g_2'-g_1'g_2$ is the Wronskian of the auxiliary equation [40]. One can also see that $g/\theta^{1/2}$ satisfy the Bessel equation, so that $g_1=\theta^{1/2}J_0(s)$, $g_2=\theta^{1/2}Y_0(s)$, and $W=2/\pi$, where $J_0$ and $Y_0$ are the Bessel functions, $s=b_1^{1/2}\theta$. Then, finally, we find the general solution of Eq. (9) given by (10).

When the bunch energy is sufficiently large, such that $4\theta^2=4\gamma>>1/b_1$ (corresponding to the condition $(2s)^2>>1$), one finds the following integral of Eq. (B1): $(d\rho/d\theta)^2+b_1\rho^2+b_2\rho^{-2}=C$, where constant $C$ is equal to the initial value of the expression on the left-hand-side. One has from this integral:

$$d\rho/d\theta = \pm(C-b_1\rho^2-b_2\rho^{-2})^{1/2}, \tag{B4}$$

where the $\pm$ sign corresponds to the sign of $d\rho/d\theta=\sigma/2\gamma^{1/4}-(2\gamma^{3/4}/E_z)(d\sigma/d\tau)$. If $b_1>0$ (focusing wakefield), integration in (B4) gives the solution (11).

### C: Simulation of electron bunch dynamics

The initial electron bunch injected into a laser wakefield was modeled numerically by a random Gaussian distribution (which is obtained from the standard uniform random-number generator via the Box-Muller transformation) in both longitudinal and transversal directions, with an average electron concentration given by $n_b=n_{b0}\exp[-((x-x_c)^2+(y-y_c)^2)/r_b^2-(z-z_c)^2/\sigma_b^2]$, where $(x_c, y_c, z_c)$ is the center of the bunch and $r_b$ and $\sigma_b$ are constants. The energy spread in the bunch and the emittance are determined by the random Gaussian distribution around the central values in the momentum space. The channel-guided laser wakefield is calculated by our fluid-Maxwell code [12, 30]. Then, the bunch dynamics in the wakefield is studied by numerically solving the equations (3) with the 4-th order Runge-Kutta method for each particle with corresponding initial position and momentum. The root-mean-squared (rms) values are calculated according to

$$x_{rms} = (<x^2> - <x>^2)^{1/2}, \tag{C1}$$

where $<>$ denotes averaging over all particles. The normalized rms emittance is calculated from (see, e.g., Ref. [36])

$$\varepsilon_{nx} = [x_{rms}^2 p_{x,rms}^2 - (<xp_x> - <x><p_x>)^2]^{1/2}. \tag{C2}$$

Here $x$ is the transverse coordinate, $p_x$ is the transverse momentum component normalized to $m_e c$, and $x_{rms}$ and $p_{x,rms}$ are the corresponding rms values calculated from (C1). The y-emittance $\varepsilon_{ny}$ is defined the same way. It has to be noted that the expression (C2) should be used instead of the normalized trace-space emittance when the energy spread is not very small [36]. This is the case for e-bunches generated by the laser wakefield acceleration mechanism.

FIGURE CAPTIONS

FIG. 1. (Color) The longitudinal ($E_z$) and transverse ($F_r$) components of the channel-guided laser wakefield generated by a Gaussian laser pulse. The laser pulse (not shown) is centered at $\xi=-6$ and travels to the right. In this case $\sigma=2$, $w=5$, and $a_0=0.9$.

FIG. 2. The near-axis behavior of the laser wakefield presented in Fig. 1.

FIG. 3. The transverse dynamics of an electron in the laser wakefield presented in Figs. 1 and 2. Electrons with an initial energy of 102.2 MeV ($\gamma_0=200$) are injected off-axis at different longitudinal positions, namely, at $\xi=-9.5$ (the oscillating curve with the long period), $\xi=-10.5$ and $-11.5$ (the two closely spaced curves, which grow monotonously), and at $\xi=-13.5$ (the curve with a long period).

FIG. 4. The rms radius of an electron bunch as a function of propagation distance in the laser wakefield presented in Fig. 1. The injection energy is 204.4 MeV ($<\gamma>_0=400$), the normalized transverse emittances $\varepsilon_{nx}$ and $\varepsilon_{ny}$ are 0.95 µm, the rms relative energy spread is 2%, and the FWHM bunch duration is 0.75 fs. The initial rms bunch radiuses are 0.96 µm (solid curve), 1.91 µm (dashed curve) and 2.87 µm (dashed-dotted curve). The bunches are injected into the accelerating region on-axis, at $\xi=-9.5$ (see Figs. 1 and 2).

FIG. 5. Injected (a) and accelerated (b) e-bunches in $x-z$ plane. Initially $<\gamma>_0=400$, the FWHM bunch duration is 7.5 fs, $\sigma_{x,y}=2.87$ µm, $\varepsilon_{nx,y}=0.95$ µm, rms energy spread is 2 %. The bunch is injected into the laser wakefield at $\xi=-9.5$ (see Figs. 1 and 2). The plasma channel length is 5 cm.

FIG. 6. Evolution of rms bunch radius during acceleration in the laser wakefield for the case simulated in Fig. 5.

FIG. 7. Femtosecond electron-bunch after 50 cm propagation in a vacuum drift space. The initial bunch parameters are: FWHM duration of 7.5 fs, $\sigma_{x,y}=1.91$ µm, $\varepsilon_{nx,y}=0.95$ µm, $<\gamma>=400$, rms energy spread of 2 %.

FIG. 8. Evolution of the bunch emittance and rms bunch length in vacuum for the case simulated in Fig. 7.

FIG. 9. (Color) Evolution of parameters of a femtosecond electron bunch injected off-axis into the laser wakefield, at different phases of the field. The bunch is injected at $\xi=-9.6$ (see Figs. 1 and 2; this case corresponds to the black curves), at $\xi=-9.3$ (red curves), at $\xi=-9$ (green curves), and at $\xi=-8.7$ (blue curves). Initially $<x>=9.55$ µm, $<y>=0$, mean bunch

energy is 204.4 MeV ($<\gamma>_0=400$), FWHM duration is 7.5 fs, $\sigma_{x,y}=1.43$ μm, $\varepsilon_{nx,y}=0.95$ μm, and the rms energy spread is 2 %.

FIG. 10. Electron bunch after acceleration in a 5 cm plasma channel for off-axis injection. The bunch is injected at $\xi=-9.6$ (see Figs. 1 and 2), the other parameters of the calculation are the same as in Fig. 9.

FIG. 11. Scheme of a two-stage laser wakefield accelerator with a short distance between the stages. A first laser pulse (depicted by the solid arrows) is tightly focused to a gas jet (grey ellipse) and produces a relativistic femtosecond electron-bunch (black ellipse) in the bubble regime. The bunch is injected into a channel-guided laser wakefield driven by a second broader laser-pulse (depicted by dashed arrows) and then further accelerated in the channel to ultra-relativistic energies.

TABLE I. Transverse size and the emittance of the femtosecond electron-bunch, after 50 cm propagation in vacuum, depending on the mean energy and the charge of the bunch. The initial bunch parameters are the same as for Fig. 7.

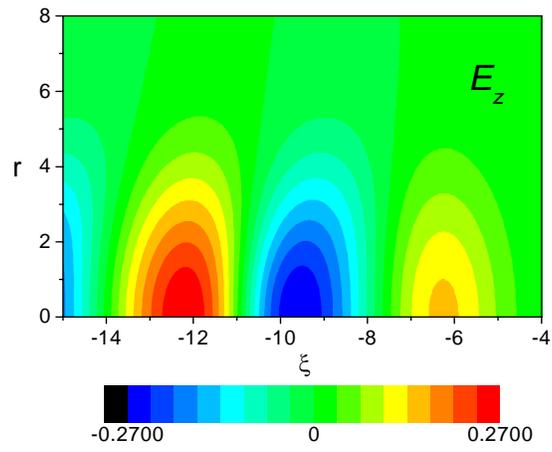

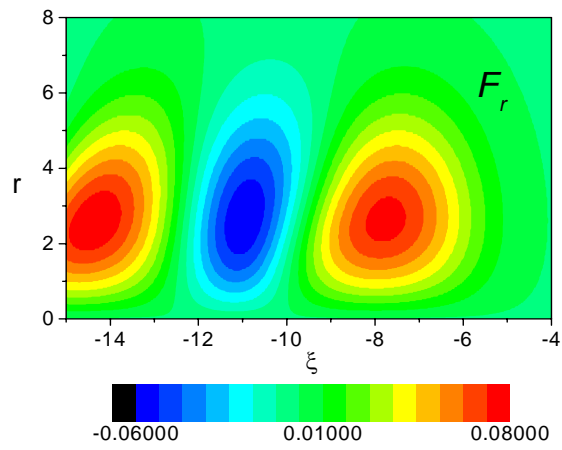

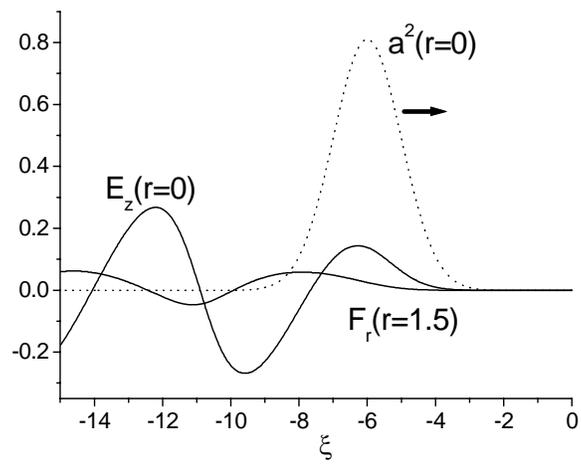

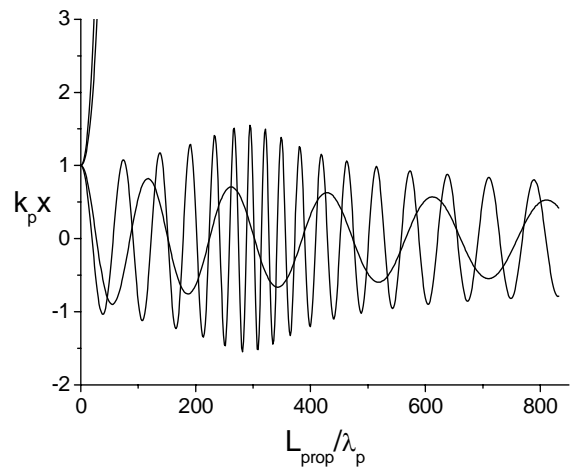

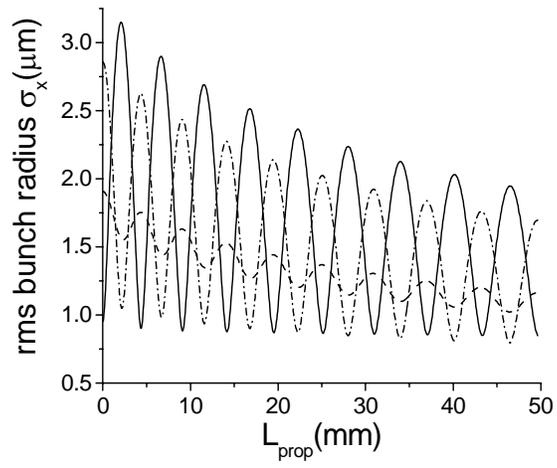

FIG. 5
A.G. Khachatryan

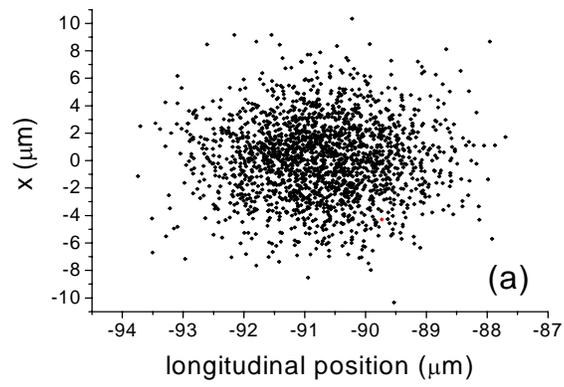

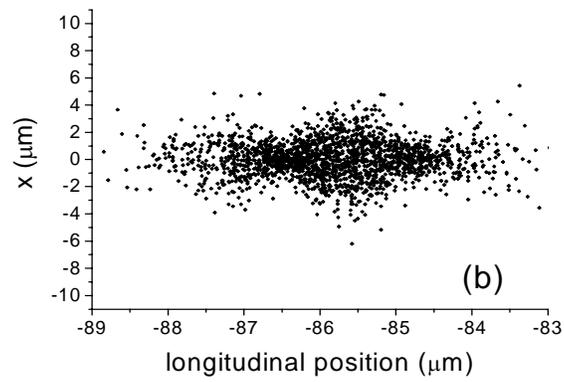

FIG. 6
A.G. Khachatryan

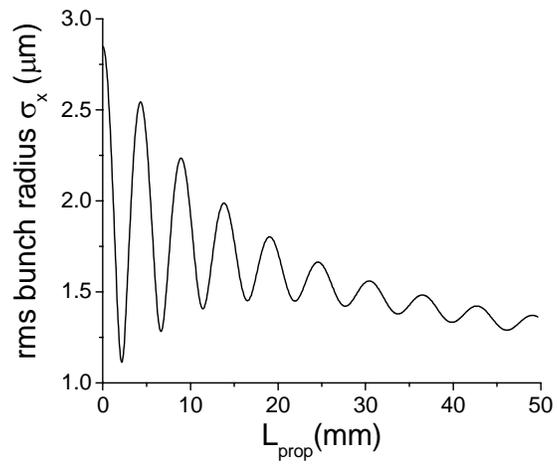

FIG. 7
A.G. Khachatryan

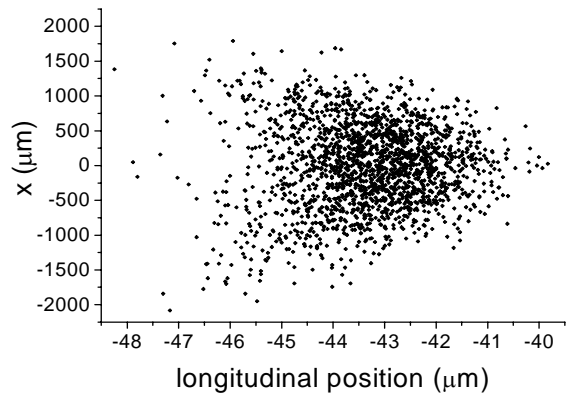

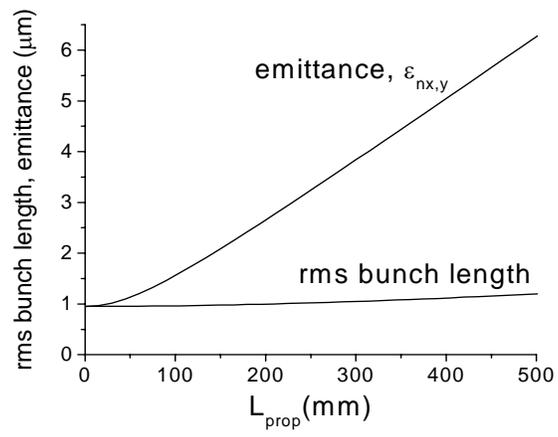

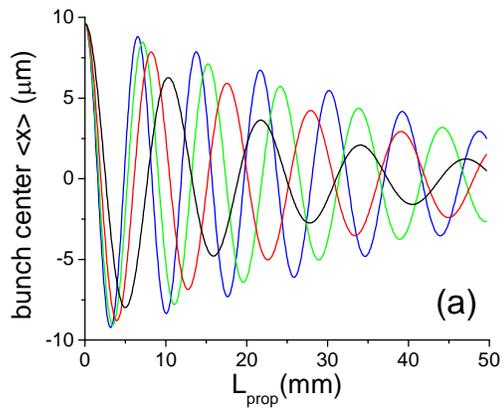
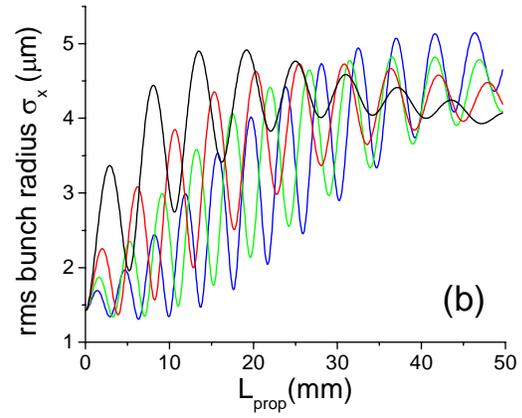
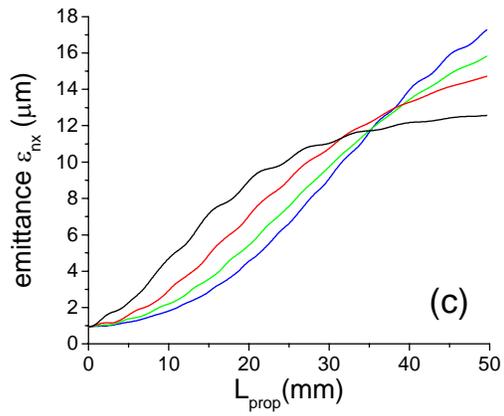
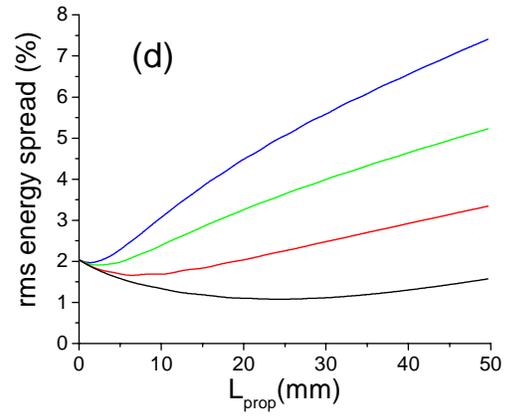

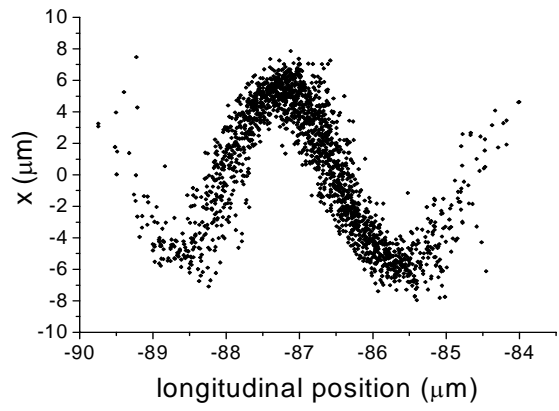

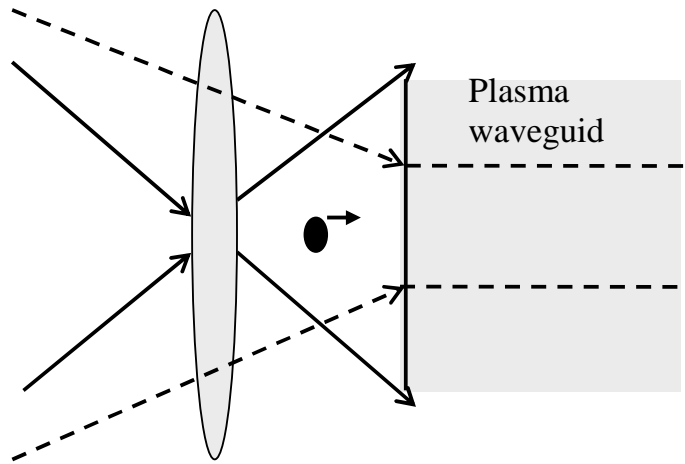

Table I.
A.G. Khachatryan

|            | $\langle\gamma\rangle_0=200$ |                       | $\langle\gamma\rangle_0=400$ |                       | $\langle\gamma\rangle_0=800$ |                       |
|------------|------------------------------|-----------------------|------------------------------|-----------------------|------------------------------|-----------------------|
| Charge(pC) | $\sigma_{x,y}$ (μm)          | $\varepsilon_{nx,y}$ (μm) | $\sigma_{x,y}$ (μm)        | $\varepsilon_{nx,y}$ (μm) | $\sigma_{x,y}$ (μm)      | $\varepsilon_{nx,y}$ (μm) |
| 0          | 1230                         | 12.3                  | 617                          | 6.2                   | 308.4                        | 3.21                  |
| 50         | 1250                         | 12.9                  | 620                          | 6.3                   | 309.1                        | 3.22                  |
| 100        | 1260                         | 14.3                  | 623                          | 6.5                   | 309.8                        | 3.24                  |
| 200        | 1320                         | 18.7                  | 629                          | 7.1                   | 311.2                        | 3.3                   |
| 500        | 1340                         | 39.2                  | 645                          | 10.7                  | 315.1                        | 3.8                   |